\documentclass[a4paper,11pt]{article}
\pdfoutput=1

\usepackage{jheppub}  
\usepackage[T1]{fontenc} 
\usepackage{graphicx,amsmath,bm,array}
\usepackage[linktocpage=true,colorlinks=true,linkcolor=blue,citecolor=blue]{hyperref}

\usepackage{slashed}
\usepackage{lmodern}
\usepackage {amssymb}

\title{\boldmath HQ collisional energy loss in a magnetized medium}
\author[a,b,1]{Balbeer Singh,\note{Corresponding author.}}
\author[a]{Surasree Mazumder,}
\author[a]{and Hiranmaya Mishra}
\affiliation[a]{Theory Division, Physical Research Laboratory, Navrangpura, Ahmedabad 380009, India}
\affiliation[b]{Indian Institute of Technology Gandhinagar,Gandhinagar 382355, Gujarat India}
\emailAdd{balbeer@prl.res.in}
\emailAdd{surasree@prl.res.in}
\emailAdd{hm@prl.res.in}

\abstract{ We study the effect of the magnetic field on the collisional energy loss  of heavy quark (HQ) moving in a magnetized thermal partonic medium. This is investigated in the strong field approximation where the lowest Landau level (LLL) becomes relevant. We work in the limit $g\sqrt{eB}\ll T\ll \sqrt{eB}$ which is relevant for heavy ion collisions. Effects of the magnetic field are incorporated through the resummed gluon propagator in which the dominant contribution arises from the quark loop. We also take the approximation $\sqrt{eB}\ll M$, M being the HQ mass, so that the HQ is not Landau quantized. It turns out that there are only two types of scatterings that contribute to the energy loss of HQ; the Coulomb scattering of HQ with light quarks/anti-quarks and the t-channel Compton scattering. It is observed that for a given magnetic field, the dominant contribution to the collisional energy loss arises from Compton scattering process i.e., $Qg\rightarrow Qg$. On the other hand, of the two processes, the Coulomb scattering i.e., $Qq\rightarrow Qq$ is more sensitive to the magnetic field. The net collisional energy loss is seen to increase with increase in the magnetic field. For a reasonable strength of the magnetic field, the field dependent contribution to the collisional energy loss is of the same order as to the case without magnetic field which can be important for the jet quenching phenomena in the heavy ion collision experiments. }

\begin{document} 
\maketitle
\flushbottom
\section{Introduction}
\label{introduction}
Experimental observations suggest that the Heavy Ion Collisions (HICs) create a novel state of matter consisting of deconfined light quarks and gluons, called the Quark Gluon Plasma (QGP). There are a plethora of theoretical studies investigating various properties
of QGP governed by Quantum Chromo Dynamics(QCD) at high temperature(T). More recent developments indicate production of 
a strong magnetic field, with an initial strength few times  the pion mass square, i.e., $eB \sim m_{\pi}^2$ at RHIC  and few tens of pion mass square i.e., $\sim 15 m_{\pi}^2$ at LHC  in a non-central HIC~\cite{ Kharzeev:2007jp,IJMPA24, PRC83, PRC85, PLB710, AHEP2014}. The existence of such a magnetic
field opens up new directions towards the theoretical studies of properties of QGP leading to diverse experimental consequences. The imprint, the magnetic field lays on QGP brings in some of the most important theoretical studies including Chiral Magnetic Effect(CME)
\cite{NPA803, PRD78}, Chiral Magnetic Wave leading to a charge-dependent elliptic flow~\cite{PRD83, JHEP01, PRL107, PRD83Miranssky},
transport properties of QGP in magnetic field~\cite{PRDFeng,PRLfukushima,Kurian:2018dbn}, quarkonia suppression~\cite{EPJC77,PRD97096011}, dilepton and photon production~\cite{Bandyopadhyay:2016fyd,Tuchin:2010gx}, HQ drag and diffusion coefficients~\cite{fukushima,santosh}, jet quenching~\cite{Li:2016bbh} etc. To what extent the magnetic field embosses its influence on the deconfined medium depends on several salient properties of the 
HICs at the early stage. Despite high initial strength, the magnetic field eventually decays at a significant rate. 

The initial dynamics of the system of deconfined matter plays a pivotal role in deciding the longevity of the magnetic 
field at the later stages. It can be hypothesized that the magnetic field can induce an amount of 
electrical conductivity which becomes adequate enough for the decaying magnetic field to persist~\cite{Tuchin:2010gx,AHEP2013, PRD92, PRD92Mamo}. This in turn can induce a current which opposes the rate of decrease of the magnetic field as per Lenz's Law~\cite{PRC88Tuchin, PRC93Tuchin, NPA929}. Therefore, it is imperative that the external magnetic field persists long enough so that it can impart crucial effects on various properties of the medium. It is instructive to investigate the effect of this field on the space-time evolution of the medium. Some studies  are trying to see this effect by reconstructing Hydrodynamic evolution in presence of magnetic field, i.e. Magneto hydrodynamics ~\cite{NPA_MHD, EPJC_MHD, PRC_MHD}. 

In the case of QGP, a more realistic approach of estimation of the magnetic field and
its relaxation time must include the medium effects i.e., electrical conductivity ($\sigma$) of
the medium. The phenomenological models that are used to describe QGP evolution
show that the strongly interaction system in HICs is thermalised just after the collision
($\tau\sim $ 0.5fm) where magnetic field is near its maximum value \cite{Kolb:2002ve,Tuchin:2013ie}. In a conducting
medium, magnetic field satisfies a diffusion equation with diffusion coefficient $(\sigma\mu)^{-1}$, where $\mu$ is the  magnetic permeability. With this one finds that the time scale over which magnetic field remain reasonably strong over a length scale $L$ is $\tau=L^2\sigma/4$ \cite{Tuchin:2010vs}. For the electrical conductivity $\sigma\simeq 0.04$T from Ref.\cite{Yin:2013kya} at $T=200$ MeV, the electrical conductivity $\sigma=8$MeV. This leads to the relaxation time  $\tau\sim 1$ fm for a system size of the order of $10$ fm.  This also suggest that the magnetic field is a slowly varying function of time and can remain reasonably strong for a longer period of time compared to the case of without a medium. For higher temperatures $\sigma$ will be higher \cite{Gupta:2003zh} increasing the value of $\tau$. Further, it is shown in Ref.\cite{Tuchin:2013ie} in an expanding medium the magnetic field remain somewhat constant for a longer time.

Consequently, it is of utmost importance to explore to what extent the magnetic field inside QGP affects different observables of the deconfined matter. To this end, surveying the in-medium properties of heavy quarks~\cite{PRD88, PRD89, PRD93Hattori} and quarkonia have become
quite relevant in the context of magnetic field~\cite{PRC84044908, PRD88105017, PRL113, PLB751, PRD92054014, PRD91066001, JHEP01052, EPJC77, PRD97096011, EPJC78spin,CS:2018mag,CS:2018jql,Reddy:2017pqp}. However, since the HQs are moving in real time inside the QGP, understanding and estimating the dynamical properties 
of HQs are also necessary. In this context, transport coefficients like drag and diffusion of HQ have been estimated in presence of 
a strong external magnetic field in some of the recent literatures~\cite{santosh,fukushima}. AdS/CFT has also been employed to have an estimation of the drag 
force of HQ\cite{AdSCFT}. Most of the calculations with strong magnetic field have been performed using perturbative QCD (pQCD) techniques in Leading Order (LO) of the strong coupling $\alpha_s$ in the limit $M\gg \sqrt{eB}$ so that the motion of the HQ is not directly affected by the external magnetic field. Nonetheless, the light quarks/anti-quarks are  affected by the magnetic field with the gluons remaining unaffected. The thermally equilibrated light quarks are Landau quantized. The magnetic field also affects the gluon self energy through the quark loop. Further they also affect the HQ light thermal parton scattering cross-sections. 
 
It is well known that a high energy particle created in the initial stages of the heavy ion collision loses its energy in the medium by interacting with the medium partons. This leads to the phenomena of jet quenching, which as anticipated many years ago, is one of the prominent probes of QGP~\cite{Bjorken:Fermi,Xie:2019oxg,Han:2017nfz,Arleo:2017ntr}; for a recent review see Ref.\cite{Cao:2020wlm}. Generally, there are two types of processes that contribute to the energy loss namely; radiative process~\cite{Abir:2012pu,Djordjevic:2003zk,Gyulassy:2000er} and collisional process\cite{Braaten:1991jj,Peigne:2007sd,Peigne:2008nd}. Experimental results for quenching of heavy flavors\cite{Bielcik:2005wu} were suggestive of including both radiative as well as collisional energy loss has been discussed in Ref.\cite{Peigne:2007sd,Wicks:2005gt}. In the present investigation, we focus our attention to the collisional energy loss of HQ in the background of a constant magnetic field which may be relevant for the HICs  as discussed in literature~\cite{Wicks:2005gt,Connors:2017ptx,Rohrmoser:2018fkf,Edmonds:2016gys,Coci:2017lhx}. 

The present investigation intends to estimate the HQ collisional energy loss ($-dE/dx$) in the low coupling regime and strong magnetic field.  Specifically we will consider the hierarchy in the scales i.e., $\sqrt{\alpha_s eB}\ll T\ll \sqrt{eB}$ and $\sqrt{eB}\ll M$. To do so, we first calculate the resummed gluon propagator in the strong magnetic field  background. Let us note here that only quark loop contributes to the resummed gluon propagator in the limit $eB\gg T^2$. This resummed propagator is used to estimate the collisional energy loss. In this hierarchy of scales, two types of processes contribute to the scatterings of HQ with the light thermal partons  affecting the HQ energy loss. As we shall see, the collisional energy loss increases with the magnetic field and for a given magnetic field, the field dependent contribution to the collisional energy loss could be similar in magnitude to the collisional energy loss in the absence of magnetic field. This can be important for the jet quenching in the heavy ion collisions.

The paper is organized as follows. In Sec.\ref{setup}, we standardize the mathematical notations, followed by a brief description of the real-time formalism. Further, in the real-time formalism of thermal field theory, we also discuss the fermion propagator in Sec.\ref{fprop} and resummed gluon propagator in Sec.\ref{gluonresummed} in LLL approximation . In Sec.\ref{formalism}, we discuss the 
formalism to calculate HQ energy loss i.e., $-dE/dx$ with descriptions of both the cases; (a) when HQ is interacting with light quarks (Sec.\ref{HQq}) and (b) HQ scattering with the thermal gluons (Sec.\ref{HQg}). Our findings are presented in Sec.\ref{results} with the relevant plots and the possible
explanation. In Sec.\ref{summary}, we summarize the present findings and discuss the possible outlook. In the appendices, we present the detailed calculations of the gluon self energy in a magnetized thermal medium and scattering amplitudes for the process $Q  g\rightarrow Q g$.

\section{Set-up}
\label{setup}
For the present investigation, we assume here that the magnetic field is constant and is along the $\hat{z}$ direction i.e., $\vec{B}=B\hat{z}$. In the subsequent subsection, we shall discuss the quark propagator in the real-time formalism of thermal field theory and in the presence of such a magnetic field. For this purpose we use the following notation. The notations $\parallel$ and $\perp$ represents the components parallel and perpendicular to the magnetic field of the corresponding quantities. For the metric tensor, we use
\begin{equation}
g_{\mu \nu}^{\parallel}=(1,0,0,-1)  \hspace{1cm} g_{\mu \nu}^{\perp}=(0,-1,-1,0).
\end{equation}
The parallel (i.e., $a_{\mu}^{\parallel}=g_{\mu \nu}^{\parallel}a^{\nu}$) and perpendicular (i.e., $a_{\mu}^{\perp}=g_{\mu \nu}^{\perp}a^{\nu}$) components of a four-vector $a_{\mu}$ are represented as
\begin{equation}
a_{\mu}^{\parallel}=(a_{0},0,0,-a_{3}) \hspace{1cm} a_{\mu}^{\perp}=(0,-a_{1},-a_{2},0). 
\end{equation}
The four-vector product ($a^{\mu}b_{\mu}=a\cdot b$) can be written as
\begin{equation}
a \cdot b=a_{\parallel}\cdot b_{\parallel}-a_{\perp}\cdot b_{\perp}.
\end{equation}
Similarly, both the components of square of a four-vector is
\begin{equation}
a^2_{\parallel}=a_{0}^2-a_{3}^2  \hspace{1cm} a^2_{\perp}=a_{1}^2+a_{2}^2.
\end{equation}
\section{Real-time formalism}
\label{realtime}
In this section, we first summarize the basic formulation for real time thermal field theory in the presence of magnetic field in a self contained manner that will be used to obtain the results on energy loss. In thermal field theories (TFT), due to the Kubo-Martin- Schwinger conditions, the time argument for the fields varies from 0 to $-i\beta$, $\beta$ being the inverse of the temperature. In the real time formulation of TFT, the time contour can be deformed to go from $t=0$ to $t=\infty$ infinitesimally above the real axis and then back to $t=-i\beta$ below the real axis leading to a $2\times 2$ matrix structure for the propagator. Corresponding to different structure for the propagation along the contour, one can define the following four functions e.g. for fermionic fields $\psi_\alpha(x)$, ($\alpha$=1,2,3,4) as
\begin{equation}
S^{>}(X,Y)_{\alpha \beta}=-i\langle \psi_{\alpha}(X)\bar{\psi}_{\beta}(Y)\rangle
\end{equation}
\begin{equation}
S^{<}(X,Y)_{\alpha \beta}=i\langle\bar{\psi}_{\beta}(Y)\psi_{\alpha}(X) \rangle
\end{equation}
\begin{equation}
S(X,Y)_{\alpha \beta}=-i\langle \hat{T}[\psi(X)_{\alpha} \bar{\psi}_{\beta}(Y)]\rangle
\end{equation}
\begin{equation}
S^{*}(X,Y)_{\alpha \beta}=-i\langle \hat{T}^{*}[\psi_{\alpha}(X)\bar{\psi}_{\beta}(Y)]\rangle 
\end{equation}
Similarly, for bosonic fields i.e., $A^{\mu}(x)$ ($\mu=0,1,2,3$), one can define the following propagators
\begin{equation}
D^{>}_{\mu \nu}(X,Y)=-i\langle A_{\mu}(X)A_{\nu}(Y) \rangle 
\end{equation}
\begin{equation}
D^{<}_{\mu \nu}(X,Y)=-i\langle A_{\nu}(Y)A_{\mu}(X) \rangle 
\end{equation}
\begin{equation}
D_{\mu \nu}(X,Y)=-i \langle \hat{T}[A_{\mu}(X)A_{\nu}(Y)] \rangle 
\end{equation}
\begin{equation}
D^{*}_{\mu \nu}(X,Y)=-i \langle \hat{T}^{*}[A_{\mu}(X)A_{\nu}(Y)] \rangle.
\end{equation}
In the above, $\hat{T}$ and $\hat{T^*}$ are the
time ordering and anti-time ordering operators respectively which are defined as
\begin{equation}
\hat{T}[A(X)B(Y)]=\theta(x_{0}-y_{0})A(X)B(Y)\pm \theta(y_{0}-x_{0})B(Y)A(X).
\end{equation}
\begin{equation}
\hat{T}^{*}[A(X)B(Y)]=\theta(y_{0}-x_{0})A(X)B(Y)\pm \theta(x_{0}-y_{0})B(Y)A(X).
\end{equation}
In terms of these four functions, the $2\times 2$ propagator is given by, e.g. for fermion fields
\begin{equation}
S=\begin{pmatrix}
S_{11} & S_{12}\\
S_{21} &S_{22}
\end{pmatrix}\equiv\begin{pmatrix} S &S^<\\ S^> &S^*\end{pmatrix}.
\end{equation}
Here, spinor indices in the second matrix are suppressed. The $11$ component corresponds to the conventional time ordered propagator while the $22$ component corresponds to anti time ordering as the contour ordering along the contour below the real axis is conversely ordered in time. The off diagonal components correspond to the Wightman propagators. From the definitions of the functions, it is clear that all the four functions are not independent and are related by
\begin{equation}
S_{11}+S_{22}=S_{12}+S_{21}.
\end{equation}
In the Keldysh representation, a linear combination of above causal propagators is used to define the retarded ($S_{R}$), advanced ($S_{A}$) and Feynman ($S_{F}$) propagators as
\begin{equation}
S_R=S_{11}-S_{12},\quad S_A=S_{11}-S_{21},\quad S_F=S_{11}+S_{22}.
\label{allprop}
\end{equation}
Feynman propagator  can also be obtained from the advanced and the retarded propagators as
\begin{equation}
S_{F}(K)=\bigg(\frac{1}{2}-\tilde{f}(k_0)\bigg)[S_{R}(K)-S_{A}(K)]
\end{equation}
where $\tilde{f}(k_0)$ is the distribution function of the fermions. One can invert the relations of Eq.(\ref{allprop}) to re-write the propagators in the Keldysh basis in terms of those in the RA basis as 
\begin{equation}
S_{11}(K)=\frac{1}{2}\bigg(S_{F}(K)+S_{A}(K)+S_{R}(K)\bigg),
\label{s11}
\end{equation}
\begin{equation}
S_{12}(K)=\frac{1}{2}\bigg(S_{F}(K)+S_{A}(K)-S_{R}(K)\bigg),
\label{s12}
\end{equation}
\begin{equation}
S_{21}(K)=\frac{1}{2}\bigg(S_{F}(K)-S_{A}(K)+S_{R}(K)\bigg),
\label{s21}
\end{equation}
\begin{equation}
S_{22}(K)=\frac{1}{2}\bigg(S_{F}(K)-S_{A}(K)-S_{R}(K)\bigg).
\label{s22}
\end{equation}
It may be mentioned that these formulations can also be applied to nearly equilibrium systems. For thermal equilibrium systems $\tilde{f}(k_0)$ becomes the Fermi-Dirac distribution function. In the next section we discuss the quark propagator in the background of magnetic field.
\subsection{Fermion propagator in LLL}
\label{fprop}
The retarded and advanced propagators ($S_{R},S_{A}$) of a free quark of electric charge $q_f$ and mass $m$ in the presence of magnetic field $B$  can be given as~\cite{Fukushima:2019ugr}
\begin{equation}
S_{R/A}(K)=\sum_{n=0}^{\infty}\bigg[\frac{i\Xi_{n}(K)}{K^2-m^2}\bigg]_{k_{0}\rightarrow k_0\pm i\epsilon}
\label{propq}
\end{equation}
where the retarded ($R$)/advanced ($A$) corresponds to $+i\epsilon/-i\epsilon$. The sum is over all the Landau levels (LLs) that is represented by $n$. Four momentum squared $K^2=k_{0}^2-k_{z}^2-2n|q_fB|$. All LLs except the lowest ($n=0$) are doubly degenerate. The numerator of Eq.(\ref{propq}) has the Dirac structure as~\cite{Fukushima:2019ugr}
\begin{equation}
\Xi_n(K)=(\slashed{k}_{\parallel}+m)[\mathcal{P}_{+}\Theta_{n}(\zeta)+\mathcal{P}_{-}\Theta_{n-1}(\zeta)]+\slashed{k}_{\perp}\Phi_{n-1}(\zeta)
\end{equation}
where $\zeta=2 \frac{k_{\perp}^2}{|q_fB|}$ and
\begin{equation}
\Theta_{l}(\zeta)=2 e^{-\frac{\zeta}{2}}(-1)^{l}L_{l}(\zeta) 
\label{theta}
\end{equation}
\begin{equation}
\Phi_{l}(\zeta)=4 e^{-\frac{\zeta}{2}}(-1)^{l-1}L^{1}_{l-1}(\zeta).
\label{phi}
\end{equation}
$L_{l}(\zeta)$ and $L_{l}^{1}(\zeta)$ are associated Laguerre Polynomials. In Eq.(\ref{propq}), the projection operator ($\mathcal{P}_{\pm}$) that projects the spin in the direction of magnetic field is defined as $\mathcal{P}_{\pm}=(1\pm sgn(q_fB)i\gamma^{1}\gamma^{2})/2$. Note, here that the projection operator depends on the electric charge of the quark as the spin magnetic moment depends on the charge of the quark.  As mentioned earlier,  we take the strong magnetic field limit so that only LLL is relevant. We shall further assume that, at finite temperature $\sqrt{eB}\gg T,m$  so that the dynamics of light quark is governed by the magnetic field. In the LLL approximation the associated Laguerre Polynomials $L_{-1}(\zeta)=0$ and $L_{0}(\zeta)=1$, so that Eq.(\ref{propq}) reduces to
\begin{equation}
S_{R/A}(K)=i\exp\bigg({-\frac{k_{\perp}^2}{|q_fB|}}\bigg)\frac{2 (\slashed{k}_{\parallel}+m)\mathcal{P}_{+}}{k_{\parallel}^2-m^2\pm i\epsilon k_{0}},
\label{propLLL}
\end{equation}
and the Feynman propagator can be obtained from
\begin{eqnarray}
S_F(K)&=&\bigg(\frac{1}{2}-\tilde{f}(k_0)\bigg)\bigg[S_{R}(K)-S_{A}(K)\bigg]\nonumber\\
&\equiv&\bigg(\frac{1}{2}-\tilde{f}(k_0)\bigg)\rho_{F}(K).
\label{propfn}
\end{eqnarray}
where $\tilde{f}(k_0)$ is Fermi-Dirac distribution function and $\rho_{F}(K)$ is quark spectral density. From Eq.(\ref{propLLL}), it is clear that in the limit $k_{\perp}^2\ll q_fB$, the motion of a quark is restricted in the transverse directions and allowed only in the direction parallel to the magnetic field. It can also be observed that for the infrared limit i.e., $m^2,T^2,k_{\perp}^2,k_{0}^2\ll eB$, the dimensional reduction from 3+1-dimension to 1+1-dimension takes place. This dimensional reduction in the LLL approximation suggests that the pairing dynamics of quarks occurs in 1+1 dimension and spontaneous chiral symmetry breaking occurs even at weak interaction between quarks in 3+1 dimension~\cite{Gusynin:1995nb}.   
\subsection{Resummed gluon propagator in LLL}
\label{gluonresummed}
Next we consider the resummed retarded/advanced gluon propagator in the presence of magnetic field within the LLL approximation. The resummed propagator is obtained by inserting the self energy corrections in the bare propagator and can be written as
\begin{equation}
D^{R/A}_{\mu \nu}(K)=[(D^{R/A}_{\mu \nu}(K))_{0}^{-1}+\Pi^{R/A}(K)_{\mu \nu}]^{-1}
\label{rprop}
\end{equation}
where the bare gluon propagator $(D^{R/A}_{\mu \nu}(K))_{0}$ in covariant gauge is given as 
\begin{equation}
(D^{R/A}_{\mu \nu}(K))_{0}=-\frac{P_{\mu \nu}(K)}{(k_0\pm i\epsilon)^2-k^2}+\xi \frac{K_{\mu}K_{\nu}}{((k_0\pm i\epsilon)^2-k^2)^2}.
\label{bareg}
\end{equation}
In Eq.(\ref{bareg}), the projection operator ($P_{\mu \nu}(K)$ is defined as  
\begin{equation}
P_{\mu \nu}(K)=-g_{\mu \nu}+\frac{K_{\mu} K_{\nu}}{(k_0\pm i\epsilon)^2-k^2}
\end{equation}
and $\xi$ is the gauge parameter. The retarded/advanced resummed gluon propagator depends on the retarded/advanced gluon self energy that, in general, can get contribution from both the gluon loop and the quark loop. The magnetic field does not affect the contribution from the gluon loop. However, the magnetic field modifies the quark loop contribution. The leading contribution from the gluon loop to the self energy at finite temperature $T$ is proportional to $g^2T^2$ while we shall see that the leading contribution from the quark loop at finite $T,B$ is proportional to $g^2 |q_fB|$ as given in Eq.(\ref{pir1}). Since we work in the limit $eB\gg T^2,m^2$, we shall drop the gluon loop contribution and keep quark loop contribution in the retarded/advanced gluon self energies. Taking the magnetic field in the $\hat{z}$ direction, the most general form of gluon self energy at finite $T$ and $B$ can be written in terms of seven independent tensors as
\begin{equation}
\Pi^{R/A}_{\mu \nu}(K)=\sum_{j=\parallel,\perp,T,L}\Pi^{R/A}(K)_{j}P^{j}_{\mu \nu}(K)+\Pi_{P}\frac{K_{\mu}K_{\nu}}{K^2}+\Pi_{n}n_{\mu}n_{\nu}+\Pi_{b}b_{\mu}b_{\nu}
\end{equation}
where $L,T,\parallel,\perp$ respectively are for longitudinal, transverse, parallel and perpendicular components of the gluon self energy. The four-vectors $n^{\mu}=(1,\textbf{0})$ and $b^{\mu}=(0,0,0,-1)$, break Lorentz and rotational symmetry due to thermal medium and the magnetic field respectively. Further, the projection operators are transverse to the momentum i.e., $P_{\mu}P^{\mu \nu}_j=0$. It turns out that only the four projection tensors contribute to the retarded/advanced self energies in the leading order. Explicitly these projection operators are given as
\begin{equation}
P^T_{\mu\nu}(K)=-g_{\mu\nu}+\frac{k_0}{k^2}\bigg[K_{\mu}n_{\nu}+n_{\mu}K_{\nu}\bigg]-\frac{1}{k^2}\bigg[K_{\mu}K_{\nu}+K^2n_{\mu}n_{\nu}\bigg],
\label{PT}
\end{equation}
\begin{equation}
P^L_{\mu\nu}(K)=-\frac{k_0}{k^2}\bigg[K_{\mu}n_{\nu}+n_{\mu}K_{\nu}\bigg]+\frac{1}{k^2}\bigg[\frac{k_0^2}{K^2} K_{\mu}K_{\nu}+K^2n_{\mu}n_{\nu}\bigg],
\label{PL}
\end{equation}
\begin{equation}
P^{\parallel}_{\mu \nu}(k)=-g^{\parallel}_{\mu \nu}+\frac{k^{\parallel}_{\mu} k^{\parallel}_{\nu}}{k_{\parallel}^2},
\end{equation}
\begin{equation}
P^{\perp}_{\mu \nu}(k)=-g^{\perp}_{\mu \nu}+\frac{k^{\perp}_{\mu} k^{\perp}_{\nu}}{k_{\perp}^2}.
\end{equation}
 The parallel and perpendicular components of self energy comes from the quark loop while the longitudinal and transverse components are from the gluon loop. Taking contribution from both quark and gluon loop the resummed retarded gluon propagator is given as~\cite{Hattori:2017xoo}
\begin{eqnarray}
D^{R}_{\mu\nu}(K)&=& -\frac{1}{\Delta(K)}\bigg[(K^2-\Pi^{\parallel}_R(K)-\Pi^{L}_R(K))P^{T}_{\mu\nu}(K)+(K^2-\Pi^{\parallel}_R(K)-\Pi^{L}_R(K))P^{L}_{\mu\nu}(K)\nonumber\\
&+&\Pi^{\parallel}_R(K)P^{\parallel}_{\mu\nu}(K)+D_{\perp}(K)P^{\perp}_{\mu\nu}(K)\bigg]+\xi\frac{K_{\mu} K_{\nu}}{(K^2)^2}
\label{gluonprpagator}
\end{eqnarray}
where 
\begin{eqnarray}
\Delta(K)&=&(K^2-\Pi^{T}_R(K))(K^2-\Pi^L_R(K))-\Pi^{\parallel}_R(K)\bigg[K^2-a\Pi^T_R(K)\frac{K^2}{k^2_{\parallel}}\nonumber\\
&-&\Pi^L_R(K)(1-a)\frac{k_0^2}{k^2_{\parallel}}\bigg],
\label{delta}
\end{eqnarray}
and
\begin{eqnarray}
D_{\perp}(K)&=&\frac{1}{K^2-\Pi^T_R(K)-\Pi^L_R(K)}\bigg[\Pi_{\parallel}(K)(\Pi^L_R(K)-\Pi^T_R(K))(1-a) \frac{k_0^2}{k^2_{\parallel}}\nonumber\\
&+& \Pi^{\perp}_R(K)(K^2-\Pi^L_R(K)-\Pi^{\parallel}_R(K))\bigg]
\end{eqnarray}
and $a=k_3^2/k^2$. In the LLL approximation for the propagator as in Eq.(\ref{propLLL}) lead to $\Pi_{\perp}=0$. This is due to the fact that  there is no current in the transverse direction in LLL approximation. All the expressions for the gluon self energies ($\Pi_L, \Pi_T$ and $\Pi_{\parallel}$ ) are explicitly given in the appendix(\ref{selfEn}). As may be noted, the gluon loop contribution to the gluon self energy is of the order of $\alpha_s T^2$ which can also be dropped with respect to the quark loop contribution which is of the order of $\alpha_seB$.  In this approximation, the resummed retarded gluon propagator becomes
\begin{equation}
D^R_{\mu\nu}(K)=-\frac{1}{\Delta}[(K^2-\Pi^{\parallel}_R)(P^{T}_{\mu\nu}+P^{L}_{\mu\nu})+\Pi^{\parallel}_RP^{\parallel}_{\mu\nu}+
D_{\perp}(K)P^{\perp}_{\mu\nu}]
+\xi\frac{K_{\mu}K_{\nu}}{(K^2)^2}
\label{propg}
\end{equation}
where
\begin{equation}
P^L_{\mu\nu}+P^T_{\mu\nu}=-g_{\mu\nu}+\frac{q_{\mu}q_{\nu}}{q^2}.
\end{equation}
Since $D_{\perp}(K)\sim \alpha_s^2 eB T^2\ll \Pi_{\parallel}\sim \alpha_s eB$, the third term in Eq.(\ref{propg}) may be dropped. Further as may be  observed from Eq.(\ref{delta}) $\Delta\approx K^2(K^2-\Pi_{\parallel}(K))$ with similar argument. Thus Eq.(\ref{propg}), taking appropriate $i\epsilon$ factors into account, can be approximated as
\begin{equation}
D^{R}_{\mu \nu}(K)=-\frac{P_{\mu \nu}(K)}{(k_0+i\epsilon)^2-k^2}-\frac{\Pi^{\parallel}_{R}(k_0+i\epsilon,\textbf{k})P^{\parallel}_{\mu \nu}(K)}{K^2(K^2-\Pi^{\parallel}_{R}(k_{0}+i\epsilon,\textbf{k}))}+\xi \frac{K_{\mu}K_{\nu}}{((k_0+i\epsilon)^2-k^2)^2}.
\label{rprop1}
\end{equation}
Similarly, the resummed advanced and Feynman gluon propagators can be written as
\begin{equation}
D^{A}_{\mu \nu}(K)=-\frac{P_{\mu \nu}(K)}{(k_0-i\epsilon)^2-k^2}-\frac{\Pi^{\parallel}_{A}(k_0-i\epsilon,\textbf{k})P^{\parallel}_{\mu \nu}(K)}{K^2(K^2-\Pi^{\parallel}_{A}(k_{0}-i\epsilon,\textbf{k}))}+\xi \frac{K_{\mu}K_{\nu}}{((k_0-i\epsilon)^2-k^2)^2},
\label{adprop}
\end{equation}
\begin{equation}
D^{F}_{\mu \nu}(K)=(1+2f(k_0))\bigg[D^R_{\mu \nu}(K)-D^A_{\mu \nu}(K)\bigg].
\label{feynprop}
\end{equation}
We shall use Eqs.(\ref{rprop1}), (\ref{adprop}) and (\ref{feynprop}) to estimate the HQ energy loss. 
\section{Formalism}
\label{formalism}
In this section, we shall calculate the energy loss of HQ of mass $M$, moving in a magnetized thermal medium of light quarks/anti-quarks and gluons, with momentum $\textbf{p}$ and energy $E=\sqrt{\textbf{p}^2+M^2}$. The rate of energy loss of a heavy fermion moving with velocity $\textbf{v}$ in a thermal medium  has been considered earlier in Ref.~\cite{Braaten:1991jj,Peigne:2007sd} for QED plasma and for QGP in Ref.~\cite{Peigne:2008nd}. In general the energy loss of the heavy fermion is given by
\begin{equation}
-\frac{dE}{dx}=\frac{1}{v}\int_{M}^{\infty}dE' (E-E')\frac{d\Gamma}{dE'}
\label{eloss}
\end{equation}
where $\Gamma$ is the interaction rate of heavy fermion with the medium particles.  For the collisional energy loss, we confine our attention to $2\rightarrow 2$ processes only as has been considered earlier~\cite{Peigne:2008nd} . For large momentum of HQ, however, the energy loss by radiative processes will also be important. In general, for the case of HQ moving in a QGP medium of light partons there can be two types of scatterings namely the Coulomb scattering i.e., $Qq\rightarrow Qq$ and Compton scattering i.e., $ Qg\rightarrow Qg$. As mentioned, we shall consider the limit $M\gg \sqrt{eB}\gg T$ so that HQ is not directly affected by the magnetic field and the light quarks are populated only in LLL. In the following sections we shall calculate the interaction rate and subsequently estimate the energy loss of the HQ.  
\subsection{Energy loss due to scattering with light quark: $ Q q\rightarrow  Q q$}
\label{HQq}
\begin{figure}[h]
\centering 
\includegraphics[width=.41\textwidth,origin=c,angle=0]{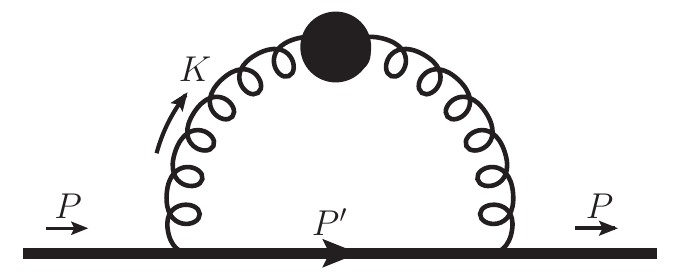}
\caption{\label{Qq} Feynman diagram for heavy quark self energy. The bold solid line refers to HQ while the curly line corresponds to gluon. The black blob here shows the gluon resummed propagator in the presence of a magnetic field background. In the LLL, the contribution to the resummed gluon propagator comes from the quark loop only.}
\end{figure}

The interaction rate $\Gamma$ of the HQ is related to imaginary part of its retarded self energy in the medium~\cite{Thoma:2000dc}
\begin{equation}
\Gamma(E)=-\frac{1}{2E}(1-\tilde{f}(E))Tr\bigg[(\slashed{P}+M)\Im\Sigma_{R}(P)\bigg]
\label{interaction}
\end{equation}
where $\tilde{f}(E)$ is Fermi-Dirac distribution function and $ \Sigma_R(P)$ is the retarded self energy of HQ as shown in Fig.(\ref{Qq}).
 
In the RTF, similar to the retarded propagator Eq.(\ref{allprop}), the retarded self energy can be written in terms of $11$ and $12$ components of self energy as
\begin{equation}
\Sigma_{R}(P)=\Sigma_{11}(P)-\Sigma_{12}(P).
\end{equation}
Using the propagators for quark and gluon in the Keldysh basis the retarded self energy of HQ can be written as
\begin{equation}
\Sigma_{R}(P)=ig^2 (t_a t_b)\int \frac{d^4K}{(2\pi)^4}\bigg[D_{11}^{\mu\nu}(K)\gamma_{\nu}S_{11}(P')\gamma_{\mu}-D_{12}^{\mu\nu}(K)\gamma_{\nu}S_{12}(P')\gamma_{\mu}\bigg].
\label{selfR}
\end{equation}
The HQ propagators are $S_{11}(K)=(\slashed{K}+M)\Delta_{11}(K)$ and $S_{12}(K)=(\slashed{K}+M)\Delta_{12}(K)$ where $\Delta_{11}(K)$ and $\Delta_{12}(K)$ are given as 
\begin{equation}
\Delta_{11}(K)=\frac{1}{K^2-M^2+i\epsilon}+2\pi i\tilde{f}(k_0) \delta(K^2-M^2)
\label{delta11}
\end{equation}
and
\begin{equation}
\Delta_{12}(K)=2\pi i\tilde{f}({k_0})\delta(K^2-M^2).
\label{delta12}
\end{equation}
Here, $\tilde{f}({k_0})$ is the fermion distribution function of the HQ. With these simplifications, Eq.(\ref{selfR}) becomes
\begin{eqnarray}
\Sigma_{R}(P)&=&ig^2 (t_a t_b)\int \frac{d^4K}{(2\pi)^4}\bigg[D_{11}^{\mu\nu}(K)\gamma_{\nu}(\slashed{P'}+M)\gamma_{\mu}\Delta_{11}(P')\nonumber\\
&-&D_{12}^{\mu\nu}(K)\gamma_{\nu}(\slashed{P'}+M)\gamma_{\mu}\Delta_{12}(P')\gamma_{\mu}\bigg].
\label{selfR1}
\end{eqnarray}
It is easier for the evaluation of the retarded self energy to convert the Keldysh propagators to the RA basis using the relations as given in Eqs.(\ref{s11}) and (\ref{s12}) for the propagators $D^{\mu \nu}_{ab}$ and $\Delta_{ab}$. Let us note that in Eq.(\ref{selfR1}), the propagators $D_{11}^{\mu \nu}$ and $D_{12}^{\mu \nu}$ are resummed gluon propagators and $S_{11},S_{12}$ for the HQ and hence the corresponding $\Delta_{11},\Delta_{12}$ are bare propagators. Eq.(\ref{selfR1}) for the retarded self energy then can be rewritten as
\begin{eqnarray}
\Sigma_R(P)&=&\frac{ig^2 (t_a t_b)}{8}\int \frac{d^4K}{(2\pi)^4}\bigg[D^{\mu\nu}_{F}(K)\lambda_{\mu\nu}\Delta_{R}(P')+D^{\mu\nu}_{A}(K)\lambda_{\mu\nu}\Delta_{R}(P')\nonumber\\
&+&D^{\mu\nu}_{R}(K)\lambda_{\mu\nu}\Delta_{F}(P')+D^{\mu\nu}_{R}(K)\lambda_{\mu\nu}\Delta_{A}(P')\bigg],
\label{sigmar}
\end{eqnarray}
where $\lambda_{\mu\nu}=\gamma_{\mu}(\slashed{P'}+M)\gamma_{\nu}$. The leading contribution to the self energy comes from the gluon propagator arising in the soft momentum transfer  limit i.e., $|\textbf{k}|\sim g\sqrt{eB}$. In the limit $E\gg\sqrt{eB}\gg T\gg |\textbf{k}|$, out of the four terms in Eq.(\ref{sigmar}) only the first term becomes dominant. This is because $D_F^{\mu \nu}$ involves gluonic distribution function. Thus Eq.(\ref{sigmar}) reduces to 
\begin{equation}
\Sigma_{R}(P)=\frac{ig^2 (t_a t_b)}{8}\int \frac{d^4K}{(2\pi)^4}\bigg[D_F^{00}\gamma_{0}(\slashed{P'}+M)\gamma_{0}+2 D_F^{0i}\gamma_{0}(\slashed{P'}+M)\gamma_{i}+D_F^{ij}\gamma_{i}(\slashed{P'}+M)\gamma_{j}\bigg]\Delta_{R}(P').
\label{selfE}
\end{equation}
Using Eqs.(\ref{rprop1}),(\ref{adprop}) and (\ref{feynprop}) in the Eqs.(\ref{interaction}) and (\ref{sigmar}), the interaction rate of HQ can be given as
\begin{eqnarray}
\Gamma(E)&=&\frac{g^2}{8 E}(1-\tilde{f}(E))\int\frac{d^4K}{(2\pi)^4}(1+2f(k_0))Tr\bigg[(\slashed{P}+M)\gamma_{0}(\slashed{P'}+M)\gamma_{0}P_{\parallel}^{00}\nonumber\\
&+&2(\slashed{P}+M)\gamma_{0}(\slashed{P'}+M)\gamma_{i}P_{\parallel}^{0j}+(\slashed{P}+M)\gamma_{i}(\slashed{P'}+M)\gamma_{j}P_{\parallel}^{ij} \bigg]\rho_L\Im(\Delta_R(P'))
\end{eqnarray}
where 
\begin{equation}
\rho_L=\frac{2 \Im \Pi_{R}^{\parallel}(K)}{(K^2-\Re\Pi_{R}^{\parallel}(K))^2+(\Im \Pi_{R}^{\parallel}(K))^2}.
\label{rho}
\end{equation}
In the above equation, $\Im \Pi^{R}_{\parallel}(K)$ and $\Re\Pi^{R}_{\parallel}(K)$ respectively are the imaginary and the real parts of retarded gluon self energy as defined in Eqs.(\ref{pir1}) and (\ref{a13}). Further, let us note that the imaginary part in the retarded self energy of HQ comes from the propagator $\Delta_{R}(P')$ which can be written as
\begin{equation}
\Im(\Delta(P'))=-\frac{\pi}{2E'}\bigg(\delta(E-k_0-E')+\delta(E-k_0+E')\bigg).
\label{imgdelta}
\end{equation}
The second delta function in Eq.(\ref{imgdelta}) does not contribution due to kinematic reasons ($k_0\ll M$). Therefore, we drop this term and continue with the first delta function. Thus, the interaction rate becomes 
\begin{eqnarray}
\Gamma(E)&=&-\frac{g^2}{8 E}(1-\tilde{f}(E))\int \frac{d^4K}{(2\pi)^4}(1+2f(k_0))\bigg[(4EE'+4p^2-4\textbf{p}\cdot \textbf{k}+4M^2)k_z^2+8(E(\textbf{p}'\cdot \textbf{k})\nonumber \\
&+&E'(\textbf{p}\cdot \textbf{k}'))k_0+4(\textbf{p}\cdot(\textbf{p}-\textbf{k})+E E'-M^2)k_0^2\bigg]\frac{\rho_L\pi}{2E'k_{\parallel}^2}\delta(E-k_0-E').
\label{intrate}
\end{eqnarray}
Since we are working in the limit where $|\textbf{k}|\sim g\sqrt{eB}\ll T$ so in this limit $E\gg |\textbf{k}|$ and  one can write $E'=\sqrt{(\textbf{p}'^2+m^2)}\equiv E-\textbf{v}\cdot \textbf{k}$. The delta function in the above equation can be simplified as $\delta(E-k_0+E')\sim \delta(k_0-\textbf{v}\cdot \textbf{k})$. Further, we assume that HQ moves parallel to the magnetic field. The integration over azimuthal angle in Eq.(\ref{intrate}) can be done trivially and the integration over the polar angle can be done by using the energy delta function. The energy loss is thus obtained as
\begin{eqnarray}
-\frac{dE}{dx}\bigg|_{Qq\rightarrow Qq}&=&-\frac{g^2}{2 E}(1-\tilde{f}(E))\int_{0}^{\infty}\frac{k dk}{(2\pi)^2}\int_{-kv}^{kv} \frac{k_0 dk_0}{2\pi}(1+2f(k_0))\bigg[4(2 E^2-2E k_0)k_0^2 \nonumber\\
&-&4v^2(2 E^2k_0-E k^2)k_0+4v^2(2E^2-2 M^2-2 Ek_0)k_0^2\bigg]\frac{\rho_L\pi}{2E k_0^2(v^2-1)}.
\label{qeloss}
\end{eqnarray}
In the above equation we have used $1/E\approx 1/E'$ as the momentum of the gluon is soft. The magnetic field dependence in the energy loss for $Qq\rightarrow Qq$ scattering comes through $\rho_L$ as defined in Eq.(\ref{rho}) which depends on the real and the imaginary parts of retarded self energy of gluon. 
\subsection{HQ gluon Scattering: $ Q g\rightarrow  Q g$}
\label{HQg}
\begin{figure}[h]
\centering 
\includegraphics[width=.28\textwidth,origin=c,angle=0]{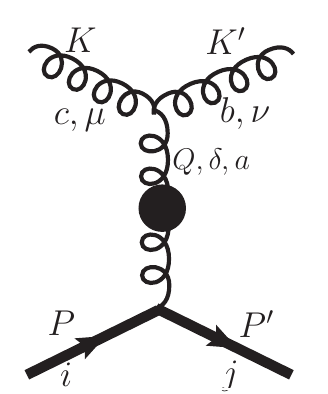}
\caption{\label{Qg} Feynman diagram for heavy quark and gluon scattering. In the strong magnetic field limit the contribution to the resummed gluon propagator comes from the quark loop only. }
\end{figure}

The other contribution to the HQ energy loss comes from the scattering of the gluons off the HQ i.e., $Q(P)+g(K)\rightarrow Q(P')+g(K')$ as shown in Fig.(\ref{Qg}). We shall discuss this process in some detail.  
The interaction rate is given as
\begin{eqnarray}
\Gamma(E)&=&\frac{1}{2E}\int\frac{d\textbf{p}'}{(2\pi)^32E'}\int\frac{d\textbf{k}}{(2\pi)^32k}f(k)
\int\frac{d\textbf{k}'}{(2\pi)^32k'}(1+f(k'))\nonumber\\
&\times&(2\pi)^4\delta^4(P+K-P'-K')|\mathcal{\bar{M}}|^2
\label{rateHQg}
\end{eqnarray}
where $|\mathcal{\bar{M}}|^2$ is the matrix element squared averaged over initial spin and color degrees of freedom and summed over final spin and color for the $Qg\rightarrow Qg$ scattering and $f(k)$ is the gluonic thermal distribution function. Generally, there are three types of processes i.e., through $s, t$ and $u$ channels that can contribute to the total scattering amplitude for this process. However, we are considering the strong magnetic field limit (LLL approximation) i.e., $M\gg \sqrt{eB}\gg T$ and assume that the effect of the  magnetic field on HQ is suppressed due to its large mass so that the $s$ and the $u$ channels, where HQ propagator arises, does not give any additional contribution arising from the magnetic field.
Therefore, the only contribution to Eq.(\ref{rateHQg}) comes from the $t$-channel scattering as shown in Fig.(\ref{Qg}). Let us note that as mentioned earlier the resummed gluon propagator has thermal contribution ($\sim \alpha_sT^2$) from gluon loop  and the magnetic field contribution ($\sim \alpha_s{eB}$) from quark loop in the gluon self energy. In LLL approximation, the thermal contributions are negligible as compared with that of the magnetic field. Consequently, we shall use the resummed gluon propagator as given in Eq.(\ref{rprop1}). The $t-$channel scattering amplitude can be written as
\begin{eqnarray}
\mathcal{M}&=&-ig^2f_{acb}t^a_{ji}[\bar{u}_j(P')\gamma^{\alpha}u_i(P)]D^{R}_{\delta\alpha}(Q)C^{\delta\mu\nu}(Q,K,-K')\epsilon_{\mu}(K)\epsilon^{*}_{\nu}(K')
\label{M}
\end{eqnarray}
where $Q=P-P'=K'-K$ is four-momentum vector for the exchanged gluon and we have kept explicitly the color indices as in Fig.(\ref{Qg}). The resulting  matrix element squared can be written after performing the color and spin sum over the final state and averaging over the initial state as 
\begin{eqnarray}
|\mathcal{\bar{M}}|^2&=&\frac{1}{4}g^4C_{F}Tr[(\slashed{P'}+M)\gamma^{\alpha}(\slashed{P}+M)\gamma^{\alpha'}]D^R_{\delta\alpha}(Q)D^R_{\delta'\alpha'}(Q)
C^{\delta\mu\nu}(Q,K,K')\nonumber\\
&\times&C^{\delta'\mu'\nu'}(Q,K,K')
\sum\epsilon_{\mu}(K)\epsilon_{\mu'}(K)\sum\epsilon_{\nu}(K')\epsilon_{\nu'}(K')\nonumber\\
&=&\frac{1}{4}g^4C_{F}Tr[(\slashed{P'}+M)\gamma^{\alpha}(\slashed{P}+M)\gamma^{\alpha'}]D^R_{\delta\alpha}(Q)D^R_{\delta'\alpha'}(Q)\nonumber\\
&\times&C^{\delta\mu\nu}(Q,K,-K')C^{\delta'}_{\mu\nu}(Q,K,-K')
\label{Msquared}
\end{eqnarray}
where $C_F=1/2$ is the color factor and $C^{\mu \nu \alpha}(P,Q,R)=(P-Q)^{\alpha}g^{\mu \nu}+(Q-R)^{\mu}g^{\nu \alpha}+(R-P)^{\nu}g^{\mu \alpha}$. To further simplify Eq.(\ref{Msquared}), we will use the transversality condition $\epsilon(K)\cdot K=\epsilon(K')\cdot K'=0$ for the gluons to obtain
\begin{eqnarray}
C^{\delta\mu\nu}(Q,K,-K')&=&-2g^{\delta\mu}K^{\nu}+g^{\mu\nu}(K+K')^{\delta}-2g^{\nu\delta}K'^{\mu}\nonumber\\
C^{\delta'}_{\mu\nu}(Q,K,-K')&=&-2g^{\delta'}_{\mu}K_{\nu}+g_{\mu\nu}(K+K')^{\delta'}-2g^{\delta'}_{\nu}K'_{\mu}
\end{eqnarray}
so that,  
\begin{equation}
C^{\delta\mu\nu}(Q,K,-K')C^{\delta'}_{\mu\nu}(Q,K,-K')=4(K^{\delta}K'^{\delta'}+K'^{\delta}K^{\delta'}). 
\label{gluon vertex}
\end{equation}
The product of propagators that appear in Eq.(\ref{Msquared}) can be simplified to 
\begin{eqnarray}
D^R_{\delta\alpha}(Q)D^R_{\delta'\alpha'}(Q)&=&\frac{g_{\delta\alpha}g_{\delta'\alpha'}}{Q^4}-\frac{\Pi^{\parallel}_R(Q)g_{\delta\alpha}P^{\parallel}_{\delta'\alpha'}+\Pi^{\parallel}_R(Q)g_{\delta'\alpha'}P^{\parallel}_{\delta\alpha}}{Q^4(Q^2-\Pi^{\parallel}_R(Q))}\nonumber\\
&+&\frac{(\Pi^{\parallel}_R(Q))^2P^{\parallel}_{\delta\alpha}P^{\parallel}_{\delta'\alpha'}}{Q^4(Q^2-\Pi^{\parallel}_R(Q))^2}.
\label{gluon propagator}
\end{eqnarray}
The first term in Eq.(\ref{gluon propagator}) corresponds to the vacuum contribution. Since we are interested in the medium contribution (i.e., $T$ and $B$), we will not consider this term. The medium dependent term that appears in Eq.(\ref{Msquared}) can be written in compact manner as
\begin{equation}
(D^R_{\delta\alpha}(q)D^R_{\delta'\alpha'}(q))(C^{\delta\mu\nu}(K,K')C^{\delta'}_{\mu\nu}(K,K'))=\mathcal{A}_{\alpha\alpha'}+\mathcal{B}_{\alpha\alpha'}+\mathcal{C}_{\alpha\alpha'}
\label{gluonvertexpropagator}
\end{equation}
where
\begin{eqnarray}
\mathcal{A}_{\alpha\alpha'}&=&-\frac{4\Pi^{\parallel}_{R}(q)}{q^4(q^2-\Pi^{\parallel}_{R}(q))}(K^{\delta}K'^{\delta'}+K'^{\delta}K^{\delta'})
g_{\delta\alpha}P^{\parallel}_{\delta'\alpha'}
\label{A1}
\end{eqnarray}
\begin{eqnarray}
\mathcal{B}_{\alpha\alpha'}&=&-\frac{4\Pi^{\parallel}_{R}(q)}{q^4(q^2-\Pi^{\parallel}_{R}(q))}(K^{\delta}K'^{\delta'}+K'^{\delta}K^{\delta'})
g_{\delta'\alpha'}P^{\parallel}_{\delta\alpha}
\label{A2}
\end{eqnarray}
\begin{eqnarray}
\mathcal{C}_{\alpha\alpha'}&=&\frac{4(\Pi^{\parallel}_{R}(q))^2}{q^4(q^2-\Pi^{\parallel}_{R}(q))^2}(K^{\delta}K'^{\delta'}+K'^{\delta}K^{\delta'})P^{\parallel}_{\delta\alpha}
P^{\parallel}_{\delta'\alpha'}
\label{A3}
\end{eqnarray}
So Eq.(\ref{Msquared}), can be written as
\begin{equation}
|\mathcal{\bar{M}}|^2=\frac{1}{4}\times\frac{1}{2}g^4(\mathcal{T}_1^{\alpha\alpha'}+\mathcal{T}_2^{\alpha\alpha'})(\mathcal{A}_{\alpha\alpha'}+\mathcal{B}_{\alpha\alpha'}+\mathcal{C}_{\alpha\alpha'})
\label{Msquared2}
\end{equation}
where $\mathcal{T}_1^{\alpha\alpha'}=Tr[\slashed{P'}\gamma^{\alpha}\slashed{P}\gamma^{\alpha'}]$ and $\mathcal{T}_2^{\alpha\alpha'}=M^2Tr[\gamma^{\alpha}\gamma^{\alpha'}]$ are traces over Dirac space. Six tensor contracted terms of Eq.(\ref{Msquared2}) are simplified in appendix(\ref{contraction}). Further simplification leads to the final form of scattering amplitude as
\begin{eqnarray}
|\mathcal{\bar{M}}|^2&=&4g^4\frac{(\Pi^{\parallel}_R(Q))^2}{Q^4(Q^2-\Pi^{\parallel}_R(Q))^2}[(P.P_{\parallel}.K)(P'.P_{\parallel}.K')+(P.P_{\parallel}.K')(K.P_{\parallel}.P')\nonumber\\
&+&(P.P')(K.P_{\parallel}.K')]-\frac{4g^4\Pi^{\parallel}_R(Q)}{Q^4(Q^2-\Pi^{\parallel}_R(Q))}[(P.K)(P'.P_{\parallel}.K')+(P.K')(K.P_{\parallel}.P')\nonumber\\
&+&(K.P')(P.P_{\parallel}.K')+(P'.K')(P.P_{\parallel}.K)-2(P.P')(K.P_{\parallel}.K')]\nonumber\\
&-&4g^4M^2\bigg[\frac{2\Pi^{\parallel}_R(Q)}{Q^4(Q^2-\Pi^{\parallel}_R(Q))}(K.P_{\parallel}.K')+\frac{(\Pi^{\parallel}_R(Q))^2}{Q^4(Q^2-\Pi^{\parallel}_R(Q))^2}(K.P_{\parallel}.K')\bigg].
\label{Msquaredfinal}
\end{eqnarray}
where the tensor product is
\begin{eqnarray}
P.P_{\parallel}.K&=&P_{\mu}P^{\mu\nu}_{\parallel}K_{\nu}=\frac{(P.q_{\parallel})(K.q_{\parallel})}{q^2_{\parallel}}-(P.k_{\parallel})
\end{eqnarray}
The expression for the retarded self energy $\Pi^{\parallel}_{R}$ is given explicitly in appendix (\ref{selfEn}). This completely defines the matrix element squared.  
\subsection*{Energy loss of HQ due to thermal gluons}
The contribution of $t-$channel Compton scattering i.e, $Qg\rightarrow Qg$, to the HQ energy loss can be obtained by using Eq.(\ref{eloss}) and the interaction rate ($\Gamma$) as given in Eq.(\ref{rateHQg}). Hence, one can write~\cite{Braaten:1991jj}
\begin{eqnarray}
-\frac{dE}{dx}\bigg|_{Qg\rightarrow Qg}&=&\frac{1}{2vE}\int\frac{d\textbf{p}'}{(2\pi)^32E'}\int\frac{d\textbf{k}}{(2\pi)^32k}f({k}_0) \int\frac{d\textbf{k}'}{(2\pi)^32k'}(1+f({k'}_0))\nonumber\\
&\times&(2\pi)^4\delta^4(P+K-P'-K')(E-E')|\mathcal{\bar{M}}|^2
\label{dEdxHQgluonic}
\end{eqnarray}
The energy and momentum transfer in each scattering are, $\omega=E-E'=|\textbf{k}'|-|\textbf{k}|$ and $\textbf{q}=\textbf{p}-\textbf{p}'=\textbf{k}'-\textbf{k}$. A comment regarding the use of resummed gluon propagator may be relevant here. The hard contributions to the energy loss ($|\textbf{q}|\sim T$) can be obtained by using the bare gluon propagator for the $Qg\rightarrow Qg$ scatterings since the self energy corrections are negligible at leading order (LO). We confine our attention here for the soft momentum transfer in the range $g\sqrt{eB}\leq |\textbf{q}|\ll T\ll \sqrt{eB}$~\cite{fukushima}. This requires the resummation of the gluon propagator  in the LLL approximation as we have used here. The $|\textbf{p}'|$ integration in Eq.(\ref{dEdxHQgluonic}) can be performed with the help of momentum delta function. In the soft momentum transfer limit the energy  $E'\approx E-\textbf{v}\cdot\textbf{q}$ so that the energy delta function reduces to  $\delta(\omega-\textbf{v}.\textbf{q})$. With these simplifications, Eq.(\ref{dEdxHQgluonic}) becomes
\begin{eqnarray}
-\frac{dE}{dx}\bigg|_{Qg\rightarrow Qg}&=&\frac{(2\pi)}{16vE^2}\int\frac{d\textbf{k}}{(2\pi)^3k}f(k)\int\frac{d\textbf{k}'}{(2\pi)^3k'}
(1+f(k'))\delta(\omega-\textbf{v}\cdot\textbf{q})\omega|\mathcal{\bar{M}}|^2.
\label{elossg}
\end{eqnarray}
Now, the simplification of the above 6-dimensional integration requires a proper choice of the co-ordinate system which should also be compatible with the terms appearing in the matrix element squared. For this purpose, we choose the direction of the momentum of the incoming HQ  along the $\hat{z}$-axis
which is also the direction of the magnetic field. We denote the angle made by $\textbf{k}$ and $\textbf{k}'$ with the $\hat{z}$-axis as $\theta_k$ and $\theta_{k'}$
respectively. Further, the azimuthal angles made by the two momenta of the incoming and outgoing gluon are $\phi_k$ and 
$\phi{k'}$ respectively. Therefore, the energy loss can be written as
\begin{eqnarray}
-\frac{dE}{dx}\bigg|_{Qg\rightarrow Qg}&=&\frac{1}{(2\pi)^516vE^2}\int kdkf(k)\int k'dk'(1+f(k'))\int^{1}_{-1}dx\int^{1}_{-1}dy\int^{2\pi}_{0}d\phi_k\nonumber\\
&\times& \int^{2\pi}_{0}d\phi_{k'}|\mathcal{\bar{M}}|^2\omega\delta(\omega-vk'y+vkx)
\label{elosscoord}
\end{eqnarray}
where, $x=\cos\theta_k$ and $y=\cos\theta_{k'}$. Now, we introduce a new integration variable $\omega$ which will take care of one of the angular integrations $x$ or $y$.
\begin{equation}
\int^{\infty}_{0}d\omega \delta[\omega-k(1-vx)]=1
\end{equation}
Using these two $\delta$-functions, we perform $x$ and $y$ angular integrations by writing the $\delta$-functions as 
$\delta[\omega-k(1-vx)]=\frac{1}{vk}\delta(x-\frac{k-\omega}{vk})$ and $\delta(\omega-vk'y+vkx)=\frac{1}{vk'}\delta(y-\frac{\omega+vkx}{vk'})$.
So, the final expression of the energy loss becomes:
\begin{eqnarray}
-\frac{dE}{dx}\bigg|_{Qg\rightarrow Qg}&=&\frac{1}{(2\pi)^516vE^2}\int^{\infty}_{0}\omega d\omega\int^{\frac{\omega}{1-v}}_{\frac{\omega}{1+v}}f(k)dk
\int(1+f(k'))dk'\int^{2\pi}_{0}d\phi_k\nonumber\\
&\times& \int^{2\pi}_{0}d\phi_{k'}|\mathcal{\bar{M}}|^2
\label{geloss}
\end{eqnarray}
The explicit form of the four-vector product and the tensor contractions in the matrix element squared are elaborately given in the appendix(\ref{ac}).

\section{Results and discussion}
\label{results}
\begin{figure}[h]
\centering 
\includegraphics[width=.44\textwidth,origin=c,angle=0]{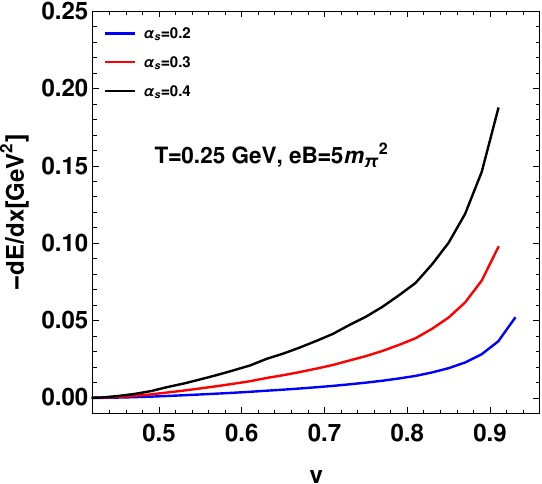}
\includegraphics[width=.46\textwidth,origin=c,angle=0]{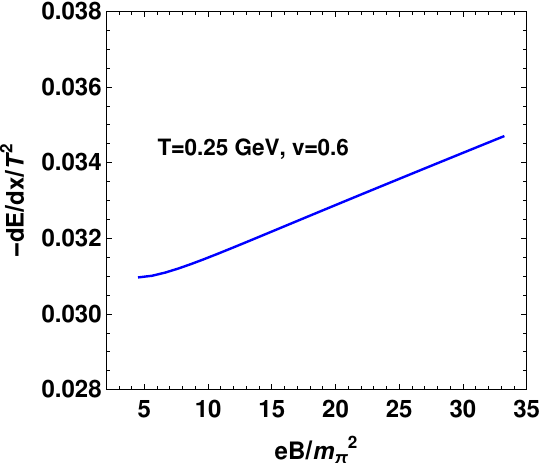}
\caption{\label{eloss1} \textbf{Left panel:} Magnetic field dependent contribution to the collisional energy loss as function of HQ velocity. Here we have taken the light quark mass $m_f=10$ MeV, flavor $n_f=2$, magnetic field $eB=0.1$ GeV$^2$($\sim 5m_{\pi}^2$) and temperature $T=0.25$ GeV. The blue, red and black curves corresponds to the coupling constant $\alpha_s=0.2$, $\alpha_s=0.3$ and $\alpha_s=0.4$ respectively. \textbf{Right panel:} Variation of the collisional energy loss scaled with $T^2$ as a function of scaled variable $eB/m_{\pi}^2$ for light quark mass $m_f=10$ MeV,  temperature $T=0.25$ GeV and coupling constant $\alpha_s=0.2$.  }
\end{figure}

The two scatterings that contribute to the HQ energy loss are $Qq\rightarrow Qq$ evaluated in Eq.{\ref{qeloss}} and $t$ channel $Qg\rightarrow Qg$ evaluated in Eq.{\ref{geloss}}. as mentioned earlier, we have confined our attention to the case where the typical momentum transfer from light partons to HQ is soft i.e., $g\sqrt{eB}\leq |\textbf{k}|\ll T \ll M$. Therefore, we have used the resummed gluon propagator in the evaluation of the matrix element squared for these two processes. To simplify our calculations for energy loss we take HQ momentum $\vec{p}=(0,0,p_z)$ and magnetic field $\vec{B}=B\hat{z}$. For the numerical purpose we take HQ as the charm quark with mass $M=1.2$ GeV, temperature $T=0.25$ GeV and magnetic field $eB=0.1$ GeV$^2$ ($\sim 5m_{\pi}^2$) so that the condition $eB\gg T^2$ is satisfied. Since $eB\gg T^2$, only LLL will be populated by the light quarks. We also take finite mass ($m_f$) of light quarks in the gluon self energy diagram. The energy loss contributions using Eqs.(\ref{geloss}) and (\ref{qeloss}) has been plotted in Fig.(\ref{eloss1}). In the left side of Fig.(\ref{eloss1}), we have shown the variation of the energy loss taking contributions from both the scattering processes as a function of velocity of HQ. We have taken here $m_f=10$ MeV for the light quark mass and number of flavors $n_f=2$. At low velocity the collisional energy loss is very small $\sim 10^{-6}$ GeV$^2$, however with increase is HQ velocity the energy loss increases.
While both the scattering processes contribute to the energy loss, it is observed numerically that the $Qg\rightarrow Qg$ gives the dominant contribution to the energy loss for a given value of velocity of the HQ essentially due to larger color factor compared to the Coulomb scattering.  

 This behavior is similar to the case of vanishing magnetic field.  It ought to be mentioned here that in the presence of magnetic field the energy loss is of similar order as compared to the case of vanishing magnetic field as estimated in Ref.{\cite{Peigne:2008nd}}. Indeed, the asymptotic value for the energy loss ($v\rightarrow 1$) of Ref.{\cite{Peigne:2008nd}} is given by
\begin{equation}
\frac{dE}{dx}=\frac{4 \pi}{33-2n_f}m_D^2
\end{equation}
with $m_D^2=4\pi\bigg(1+\frac{n_f}{6}\bigg)\alpha_s T^2$ which with the parameters $\alpha_s=0.3, n_f=2$ and $T=0.25$ GeV turns out to be $0.15$ GeV$^2$. 
This may be compared with the magnetic field contribution given by the red curve in the left panel of Fig.(\ref{eloss1}) which reaches to the value $0.1$ GeV$^2$ in the same limit. Further, it may be relevant to  note that in the limit of vanishing  light quark mass i.e., $m_f=0$, the magnetic field contribution to the gluon self energy vanishes as $\Pi_{\parallel}^{R}$ is proportional to $m_f^2$. This will lead to vanishing of the magnetic field dependent contribution to the energy loss. This is similar to the vanishing of magnetic field contribution for the dilepton production~\cite{Bandyopadhyay:2016fyd} in the same limit. 

In the right panel of Fig.(\ref{eloss1}), we display the energy loss scaled with $T^2$ as a function  of magnetic field scaled by pion mass square i.e., $eB/m_{\pi}^2$, where $m_{\pi}$ is pion mass. We have taken here $T=0.25$ GeV and the magnitude of the heavy quark velocity $v=0.6$. As may be  observed in the figure the energy loss increases with the increase in the magnetic field. Numerically, it  is seen that $Qg\rightarrow Qg$ contribution is not affected too much with the magnetic field. For this process, the magnetic field dependence arises from the resummed gluon propagator with the field dependent contribution of the quark loop. This quark loop contribution increases with the magnetic field leading to a mild decrease of the energy loss due to this process as quark loop contribution lies in the retarded propagator.  On the other hand, for the Coulomb scattering process i.e., $Qq\rightarrow Qq$ the contribution increase with increase in the magnetic field. This can be understood as follows; The energy loss is proportional to  $\rho_L$ that depends on real and imaginary parts of retarded self energies and related to the spectral function. The spectral function increases with increase in the magnetic field. This increase of $\rho_L$ with the magnetic field was also observed in Ref.\cite{Bandyopadhyay:2016fyd}. It turns out that this increase is significant and the contribution becomes similar order as the $Qg\rightarrow Qg$ for larger magnetic field leading to increase of the total energy loss with magnetic field as observed in Fig.(\ref{eloss1}).

\section{Summary and Conclusion}
\label{summary}
In the present investigation, we have studied the effect of the magnetic field on the HQ collisional energy loss in a thermalized QGP medium. We have done the analysis for the case of strong field limit i.e., $\sqrt{eB}\gg T$ so that the light quarks are populated only in the LLL. On the other hand, the heavy quark mass $M$ is much larger than the strength of the magnetic field (i.e., $M\gg \sqrt{eB}$), so that heavy quark is not Landau quantized. The effect of the magnetic field manifests through resummed retarded/advanced gluon propagator through quarks loop. Since the gluon loop contribution to the gluon self energy is proportional to $\alpha_s T^2$ and the quark loop contribution is proportional to $\alpha_s eB$, in the limit $\sqrt{eB}\gg T$ the gluon loop contribution in the gluon resummed propagator is not taken into account. For the scattering of HQ with the thermalized light partons we have considered the soft momentum transfer limit i.e., $g\sqrt{eB}\leq |\textbf{k}|\ll T\ll \sqrt{eB}$. With $M\gg \sqrt{eB}$ and the in the LLL approximation the relevant scattering processes are Coulomb scattering i.e., $Qq\rightarrow Qq$ and $t$-channel Compton scattering $Qg\rightarrow Qg$. The $u$ and the $s$ channels of the Compton scatterings are not affected by the magnetic field. For a given magnitude of the heavy quark velocity, the Compton scattering process is dominant over the Coulomb scattering process for the range of the magnetic field considered here. 

It is observed that of the two processes, the Coulomb scattering process  is more sensitive to the magnetic field as its contribution to the energy loss is proportional to the spectral function which increases with increase in the magnetic field. This leads to a net increase in the energy loss with increase in the magnetic field. It turns out that in this strong field limit, the magnetic field dependent collisional energy loss (for $eB=5 m_{\pi}^2$) is comparable to the same in the vanishing field limit and therefore could be important for the jet quenching phenomena in HICs. However, In a realistic situation in HIC up to what extent the magnetic field can affect the collisional energy loss will also depend on the medium response to the magnetic field and require  further investigations. In the vacuum, the magnetic field decreases very rapidly however, in a system with finite electrical conductivity, magnetic field satisfies diffusion equation and the relaxation time of the external magnetic field depends on the electrical conductivity of the system. In a system with larger conductivity magnetic field can sustain and resonably be strong for a longer period of time. However, it ought to be mentioned  that this requires a proper estimation of the electrical conductivity of the medium as well as solutions of magneto hydrodynamic equations which needs further investigations. Furthermore, for smaller values of magnetic field one must include the effect of higher Landau levels.

We would like to mention here that the present investigation is a  first attempt to include the effect of magnetic field in the collisional energy loss. However, for the large momentum of heavy quark, the radiative contribution to the energy loss may also be relevant and could be affected by the magnetic field. This apart, for a moderate value of magnetic field $eB\sim T^2$ the contributions  arising from the higher Landau levels could also be important for the energy loss and hence jet quenching. In this case, one must take the contribution of gluon loops in the gluon self energy for the resummed gluon propagator as well as the contribution from the $s$ and $u$ channels of Compton scattering. Some of these problems are relegated to future investigations. We also emphasize on the fact that our results are reliable only in the strong magnetic field limit where the LLL approximation is reasonable.
 
\appendix

\section{Gluon self energy}
\label{selfEn}
In the RTF, the retarded self energy of gluon in the Keldysh basis can be written as  
\begin{equation}
\Pi^{\mu \nu}_{R}(p_0,\textbf{p})=\Pi_{11}^{\mu \nu}(p_0,\textbf{p})+\Pi_{12}^{\mu \nu}(p_0,\textbf{p})
\end{equation}
where $\Pi_{11}^{\mu \nu}$ and $\Pi_{12}^{\mu \nu}$ are $11$ and $12$ component of the gluon self energy. Both $11$ and $22$ components of self energy can be obtained by using the quark propagators to acquire
\begin{eqnarray}
\Pi_{R}^{\mu \nu}(p_0,\textbf{p})&=&-ig^2t^at^b\Omega\int\frac{d^2k_{\parallel}}{(2\pi)^2}\bigg(Tr[\gamma^{\mu}S_{11}(Q)\gamma^{\nu}S_{11}(K)-\gamma^{\mu}S_{21}(Q)\gamma^{\nu}S_{12}(Q)]\bigg).
\label{pimunu}
\end{eqnarray}
where
\begin{equation}
\Omega=\frac{|q_fB|}{8\pi}\exp\bigg({-\frac{p_{\perp}^2}{|2q_fB|}}\bigg).
\end{equation}
In Eq.(\ref{pimunu}), the $B$ dependent term (i.e., $\Omega$) comes from the transverse part in the quark propagator. In the LLL, the dynamics in the transverse direction is restricted due to dimensional reduction from (3+1)-dimension to (1+1)-dimension. Therefore, the gauge invariant form of the gluon self energy in a magnetic field background can be written as
\begin{equation}
\Pi_{R}^{\mu \nu}(p_0,\textbf{p})=\Pi^{\parallel}_{R}(P) \bigg(g_{\parallel}^{\mu \nu}-\frac{p_{\parallel}^{\mu}p_{\parallel}^{\mu}}{p_{\parallel}^2}\bigg).
\label{a4}
\end{equation} 
To estimate the energy loss, we need $\Pi^{\parallel}_{R}(P)$ which can be obtained from the relation $\Pi^{\parallel}_{R}(P)=-(p_{\parallel}^2/p_z^2)\Pi_{R}^{00}(p_0,\textbf{p})$. Therefore, we focus only on the time-like component ($\Pi^{00}_{R}$) of retarded self energy. Taking the trace over Dirac matrices in Eq.(\ref{pimunu}), we get
\begin{eqnarray}
\Pi_{R}^{00}(p_0,\textbf{p})&=&-i8g^2t^at^b\Omega\int\frac{d^2k_{\parallel}}{(2\pi)^2}(\mathcal{K}\cdot \mathcal{Q})\bigg(\Delta_{11}(Q)\Delta_{11}(K)-\Delta_{21}(Q)\Delta_{12}(K)\bigg)
\label{pir00}
\end{eqnarray} 
where $\Delta_{11}$ and $\Delta_{12}$ are defined in Eqs.(\ref{delta11}) and (\ref{delta12}) with $\mathcal{K}\cdot\mathcal{Q}=k_0q_0+k_zq_z+m^2$. Similar to $\Delta_{11}$, $\Delta_{21}$ can also be obtained from $S_{21}$. Eq.(\ref{pir00}) can further be simplified by writing the propagators $\Delta_{ij}$ in terms of the retarded, advanced and Feynman propagators similar to Eqs.(\ref{s11}), (\ref{s12}) and (\ref{s21}) to acquire
\begin{eqnarray}
\Pi_{R}^{00}(p_0,\textbf{p})&=&-i4g^2t^at^b\Omega\int\frac{d^2k_{\parallel}}{(2\pi)^2}(\mathcal{K}\cdot\mathcal{Q})\bigg(\Delta_F(Q)\Delta_{R}(K)+\Delta_A(Q)\Delta_F(K)\bigg).
\label{pir001}
\end{eqnarray}
Replacing $K\rightarrow -Q$ and using the relation $\Delta_{R}(-Q)=\Delta_{A}(Q)$, Eq.(\ref{pir001}) becomes
\begin{eqnarray}
\Pi_{R}^{00}(p_0,\textbf{p})&=&2\pi g^2t^at^b\Omega\int\frac{d^2k_{\parallel}}{(2\pi)^2}(\mathcal{K}\cdot\mathcal{Q})(1-2\tilde{f}(k_0))\delta(k_{\parallel}^2-m_f^2)\frac{1}{q_{\parallel}^2-m_f^2-i\epsilon q_0}
\label{pir002}
\end{eqnarray}
The first term of Eq.(\ref{pir002}) which is independent of the Fermi Dirac distribution function,  is the vacuum contribution to the time-like component of the gluon self energy. The time-like component of the gluon self energy can be separated into the vacuum and the thermal parts to get
\begin{equation}
\Pi_{R}^{00}(p_0,\textbf{p})=\Pi_{R}^{00}(p_0,\textbf{p})|_{vac}+\Pi_{R}^{00}(p_0,\textbf{p})|_{th}
\end{equation} 
where the vacuum term is given as~\cite{fukushima}
\begin{eqnarray}
\Pi_{R}^{00}(p_0,\textbf{p})|_{vac}&=&\frac{2 g^2t^at^b\Omega p_z^2}{p_{\parallel}^2+i\epsilon p_0}\bigg[1-\frac{4m_f^2}{\sqrt{p_{\parallel}^2(4 m_f^2-p_{\parallel}^2)}}\arctan\bigg(\frac{p_{\parallel}^2}{\sqrt{p_{\parallel}^2(4 m_f^2-p_{\parallel}^2)}}\bigg)\bigg].
\end{eqnarray}
For the thermal contribution to the time-like component of the gluon self energy, the energy integral in Eq.(\ref{pir002}) can be done by using the energy delta function to obtain
\begin{eqnarray}
\Pi_{R}^{00}(p_0,\textbf{p})|_{th}&=&-\frac{2\pi g^2t^at^b\Omega m_f^2 p_z^2}{p_{\parallel}^2}\bigg(\mathcal{J}_{0}(P)+ \frac{2 p_{z}}{p_{\parallel}^2}\mathcal{J}_{1}(P)\bigg).
\label{pir}
\end{eqnarray}  
with
\begin{equation}
\mathcal{J}_{a}=\int_{-\infty}^{\infty}\frac{dk_z}{2\pi E}\tilde{f}(E)\frac{k_z^{a}}{(k_z-p_z/2)^2-p_0^2/4+p_0^2 m_f^2/p_{\parallel}^2-i p_0 \epsilon}
\end{equation}
where $a=0,1$. Using Eq.(\ref{pir}) and the relation $\Pi^{\parallel}_{R}(P)=-(p_{\parallel}^2/p_z^2)\Pi_{R}^{00}(p_0,\textbf{p})$, the parallel component of the medium dependent gluon self energy can be written as
\begin{eqnarray}
\Pi^{\parallel}_{R}(P)=2\pi g^2t^at^b\Omega m_f^2 \bigg[\mathcal{J}_{0}(P)+ \frac{2 p_{z}}{p_{\parallel}^2}\mathcal{J}_{1}(P)\bigg]
\label{pir1}
\end{eqnarray}

\textbf{Imaginary part of the gluon self energy:} The imaginary part of the retarded self energy of the gluon can be obtained from the imaginary part of the $11$ component of gluon self energy by using the relation~\cite{bellac}
\begin{equation}
\Im \Pi_{R}^{\mu \nu}(p_0,\textbf{p})=(1-\tilde{f}(p_0))\Im \Pi_{11}^{\mu \nu}(p_0,\textbf{p})
\label{a13}
\end{equation}
where 
\begin{equation}
\Pi_{11}^{\mu \nu}(p_0,\textbf{p})=ig^2 t^a t^b\int \frac{d^2{k}_{\perp}}{(2\pi)^2}e^{-\frac{k_{\perp}^2+q_{\perp}^2}{|q_fB|}}\int \frac{d^2 k_{\parallel}}{(2\pi)^2}Tr[\gamma^{\mu}S_{11}(K)\gamma^{\nu}S_{11}(Q)].
\label{pi11}
\end{equation}	
$S_{11}(P)$ in Eq.(\ref{pi11}), is the $11$ component of the quark propagator in RTF that can be obtained from the retarded, advanced and Feynman propagators by using the relation as given  in Eq.(\ref{s11}). Using the Eqs.(\ref{propLLL}) and (\ref{propfn}), the imaginary part of $\Pi_{11}^{\mu \nu}(P)$ can be written as
\begin{eqnarray}
\Im \Pi_{11}^{\mu \nu}(p_0,\textbf{p})&=&\pi g^2 t^{a}t^{b}\Omega \int_{-\infty}^{\infty} \frac{dk_{z}}{2\pi}\frac{1}{4 E_k E_q}\bigg[\bigg(1-\tilde{f}(E_k)-\tilde{f}(E_q)+2\tilde{f}(E_k)\tilde{f}(E_q)\bigg)\nonumber\\
&\times&\bigg(\mathcal{N}^{\mu \nu}(k_0=E_k)\delta(p_0-E_k-E_q)+\mathcal{N}^{\mu \nu}(k_0=-E_k)\delta(p_0+E_k+E_q)\bigg)\nonumber\\
&+&\bigg(-\tilde{f}(E_k)-\tilde{f}(E_q)+2\tilde{f}(E_k)\tilde{f}(E_q)\bigg)\bigg(\mathcal{N}^{\mu \nu}(k_0=-E_k)\delta(p_0-E_k+E_q)\nonumber\\
&+&\mathcal{N}^{\mu \nu}(k_0=E_k)\delta(p_0+E_k-E_q)\bigg)\bigg]
\label{ipi11}
\end{eqnarray}
where
\begin{equation}
\mathcal{N}^{\mu \nu}=Tr[\gamma^{\mu}\mathcal{S}_{0}(K)\gamma^{\nu}\mathcal{S}_{0}(Q)]
\end{equation}
with
\begin{equation}
\mathcal{S}_{0}(K)=(\slashed{k}_{\parallel}+m_f)(1+i\gamma^{1}\gamma^{2}).
\end{equation}
The imaginary part of the retarded self energy i.e., $\Im \Pi_{R}^{\parallel}$ can be obtained from  $\Im \Pi_{R}^{00}$ by using the general structure of the retarded self energy as given in Eq.(\ref{a4}). So only the time-like component of  $\Im \Pi_{11}^{\mu \nu}$ i,e., $\Im \Pi_{11}^{00}$ is relevant which can be obtained from Eq.(\ref{ipi11}). The momentum integration in Eq.(\ref{ipi11}) can be done by using the energy delta functions. Let us first simplify the energy delta functions and re-write those in terms of $k_z$.  By using the relation
\begin{equation}
\delta(f(x))=\sum_n \frac{\delta(x-x_n)}{\bigg|\frac{\partial f(x)}{\partial x}\bigg|_{x=x_n}}.
\end{equation}  
one can write
\begin{equation}
\delta(p_0-E_k-E_q)=\frac{\delta(k_z-k_z^{0})E_{k_z^0}E_{q_z^{0}}}{k_z^0(E_{k_z^0}+E_{q_z^0})}+\frac{\delta(k_z-k_z^{1})E_{k_z^1}E_{q_z^1}}{k_z^1(E_{k_z^1}+E_{q_z^1})},
\end{equation}
\begin{equation}
\delta(p_0+E_k-E_q)=\frac{\delta(k_z-k_z^{0})E_{k_z^0}E_{q_z^{0}}}{k_z^0(E_{k_z^0}-E_{q_z^0})}+\frac{\delta(k_z-k_z^{1})E_{k_z^1}E_{q_z^1}}{k_z^1(E_{k_z^1}-E_{q_z^1})},
\end{equation}
where
\begin{equation}
k_z^0=-\frac{p_z}{2}+\frac{1}{2|p_{\parallel}|}\sqrt{p_z^2 p_{\parallel}^2-4p_0^2m_f^2+p_{\parallel}^4},
\end{equation}
and
\begin{equation}
k_z^1=-\frac{p_z}{2}-\frac{1}{2|p_{\parallel}|}\sqrt{p_z^2 p_{\parallel}^2-4p_0^2m_f^2+ p_{\parallel}^4}.
\end{equation}
With further simplification, the imaginary part of $\Pi_{11}^{00}$ can be written as
\begin{eqnarray}
\Im \Pi_{11}^{00}(p_0,\textbf{p})&=&g^2 t^{a}t^{b}\Omega\pi 2m_f^2\bigg[\bigg(\frac{1}{k_z^0(E_{k_z^0}+E_{q_z^0})}+\frac{1}{k_z^1(E_{k_z^1}+E_{q_z^1})}\bigg)-(\tilde{f}(k_z^0)+\tilde{f}(q_z^0)\nonumber\\
&-&2 \tilde{f}(k_z^0)\tilde{f}(q_z^0))\bigg(\frac{2E_{k_z^0}}{k_z^0 p_z(2 k_z^0+p_z)}\bigg)-(\tilde{f}(k_z^1)+\tilde{f}(q_z^1)-2\tilde{f}(k_z^1)\tilde{f}(q_z^1))\nonumber\\
&\times&\bigg(\frac{2E_{k_z^1}}{k_z^1 p_z (2k_z^1+p_z)}\bigg)\bigg].
\label{a23}
\end{eqnarray}
As mentioned earlier, Eq.(\ref{a23}) can be used to obtain $\Im \Pi_{R}^{\parallel}$.
In the HTL limit, the thermal contribution to the transverse and longitudinal components of the gluon self energy that comes from the gluon loop is given as
\begin{equation}
\Pi^{L}_{R}(P)=\frac{g^2 T^2 N_c}{3}[-1+Q(x)]
\label{pil}
\end{equation}
and
\begin{equation}
\Pi^T_R(P)=\frac{g^2 T^2 N_c}{6}[x^2+(1-x^2)Q(x)]
\label{pit}
\end{equation}
where $x=\frac{p_0}{|\textbf{p}|}$. The function $Q(x)$ is given as
\begin{equation}
Q(x)=\frac{x}{2}\bigg(\ln\bigg|\frac{1+x}{1-x}\bigg|-i\pi \theta(1-x^2)\bigg).
\end{equation}
\section{$Qg\rightarrow Qg$ scattering}
\label{contraction}
The contracted terms of Eq.(\ref{Msquared2}) are 
\subsection*{Term-1:}
\begin{eqnarray}
\mathcal{T}_1^{\mu\nu}\mathcal{A}_{\mu\nu}&=&-\frac{16\Pi^{\parallel}_{R}(q)}{q^4(q^2-\Pi^{\parallel}_{R}(q))}[P^{\mu}P'^{\nu}+P'^{\mu}P^{\nu}-(P.P')g^{\mu\nu}]
[K_{\mu}K'^{\delta}P^{\parallel}_{\delta\nu}+K'_{\mu}K^{\delta}P^{\parallel}_{\delta\nu}]\nonumber\\
&=&-\frac{16\Pi^{\parallel}_{R}(q)}{q^4(q^2-\Pi^{\parallel}_{R}(q))}[(P.K)(K'.P_{\parallel}.P')+(P.K')(K.P_{\parallel}.P')\nonumber\\
&+&(K.P')(K'.P_{\parallel}.P)+(P'.K')(K.P_{\parallel}.P)-2(P.P')(K'.P_{\parallel}.K)].
\label{term1}
\end{eqnarray}
\subsection*{Term-2:}
\begin{eqnarray}
\mathcal{T}_1^{\mu\nu}\mathcal{B}_{\mu\nu}&=&-\frac{16\Pi^{\parallel}_{R}(q)}{q^4(q^2-\Pi^{\parallel}_{R}(q))}[P^{\mu}P'^{\nu}+P'^{\mu}P^{\nu}-(P.P')g^{\mu\nu}]
[K^{\delta}P^{\parallel}_{\delta\mu}K'_{\nu}+K'^{\delta}P^{\parallel}_{\delta\mu}K_{\nu}]\nonumber\\
&=&-\frac{16\Pi^{\parallel}_{R}(q)}{q^4(q^2-\Pi^{\parallel}_{R}(q))}[(P'.K')(K.P_{\parallel}.P)+(K.P')(K'.P_{\parallel}.P)\nonumber\\
&+&(P.K')(K.P_{\parallel}.P')+(P.K)(K'.P_{\parallel}.P')-2(P.P')(K.P_{\parallel}.K')].
\label{term2}
\end{eqnarray}
\subsection*{Term3:}
\begin{eqnarray}
\mathcal{T}_1^{\mu\nu}\mathcal{C}_{\mu\nu}&=&\frac{16(\Pi^{\parallel}_{R}(q))^2}{q^4(q^2-\Pi^{\parallel}_{R}(q))^2}[P^{\mu}P'^{\nu}+P'^{\mu}P^{\nu}-(P.P')g^{\mu\nu}][K^{\delta}P^{\parallel}_{\delta\mu}K'^{\delta'}P^{\parallel}_{\delta'\nu}
+K'^{\delta}P^{\parallel}_{\delta\mu}K^{\delta'}P^{\parallel}_{\delta'\nu}]\nonumber\\
&=&\frac{16(\Pi^{\parallel}_{R}(q))^2}{q^4(q^2-\Pi^{\parallel}_{R}(q))^2}[(K.P_{\parallel}.P)(K'.P_{\parallel}.P')+(K'.P_{\parallel}.P)(K.P_{\parallel}.P')\nonumber\\
&+&(K.P_{\parallel}.P')(K'.P_{\parallel}.P)+(K'.P_{\parallel}.P')(K.P_{\parallel}.P)+2(P.P')(K.P_{\parallel}.K')].
\label{term3}
\end{eqnarray}
Here, we make use of the identity:
\begin{equation}
g^{\mu\nu}P^{\parallel}_{\delta\mu}P^{\parallel}_{\delta'\nu}=-P^{\parallel}_{\delta\delta'}.
\end{equation}
\subsection*{Term-4:}
\begin{eqnarray}
\mathcal{T}_2^{\mu\nu}\mathcal{A}_{\mu\nu}&=&-\frac{16M^2\Pi^{\parallel}_{R}(q)}{q^4(q^2-\Pi^{\parallel}_{R}(q))}g^{\mu\nu}[K_{\mu}K'^{\delta}P^{\parallel}_{\delta\nu}
+K'_{\mu}K^{\delta}P^{\parallel}_{\delta\nu}]\nonumber\\
&=&-\frac{32M^2\Pi^{\parallel}_{R}(q)}{q^4(q^2-\Pi^{\parallel}_{R}(q))}(K.P_{\parallel}.K').
\label{term4}
\end{eqnarray}
\subsection*{Term-5:}
\begin{eqnarray}
\mathcal{T}_2^{\mu\nu}\mathcal{B}_{\mu\nu}&=&-\frac{16M^2\Pi^{\parallel}_{R}(q)}{q^4(q^2-\Pi^{\parallel}(q)_{R})}g^{\mu\nu}
[K^{\delta}P^{\parallel}_{\delta\mu}K'_{\nu}+K'^{\delta}P^{\parallel}_{\delta\mu}K_{\nu}]\nonumber\\
&=&-\frac{32M^2\Pi_{\parallel}(q)}{q^4(q^2-\Pi_{\parallel}(q))}(K.P_{\parallel}.K').
\label{term5}
\end{eqnarray}
\subsection*{Term-6:}
\begin{eqnarray}
\mathcal{T}_2^{\mu\nu}\mathcal{C}_{\mu\nu}&=&\frac{16M^2(\Pi^{\parallel}_{R}(q))^2}{q^4(q^2-\Pi^{\parallel}_R(q))^2}g^{\mu\nu}[K^{\delta}P^{\parallel}_{\delta\mu}K'^{\delta'}P^{\parallel}_{\delta'\nu}
+K'^{\delta}P^{\parallel}_{\delta\mu}K^{\delta'}P^{\parallel}_{\delta'\nu}]\nonumber\\
&=&-\frac{32M^2(\Pi^{\parallel}_{R}(q))^2}{q^4(q^2-\Pi^{\parallel}_{R}(q))^2}(K.P_{\parallel}.K').
\label{term6}
\end{eqnarray}

\section{Four vector product and tensor contractions}
\label{ac}
With the assumption that the HQ quark moves in the direction of the magnetic field, the four-vector products in the matrix element squared i.e., $|\mathcal{\bar{M}}|^2$ as given in Eq.(\ref{Msquaredfinal}) can be given as
\begin{eqnarray}
P.K&=&Ek_0-\textbf{p}.\textbf{k}=Ek_0-{p}{k}\cos\theta_k=Ek_0-{p}{k}x\nonumber\\
P.K'&=&Ek_0'-\textbf{p}\cdot\textbf{k}'=Ek_0'-{p}{k}'y\nonumber\\
P'.K&=&E'k_0-(\textbf{p}+\textbf{k}-\textbf{k}')\cdot\textbf{k}=Ek_0-{v}{k}{k}'y+vk^2x-pkx-k^2+\textbf{k}\cdot\textbf{k}'\nonumber\\
P'.K'&=&E'k_0'-(\textbf{p}+\textbf{k}-\textbf{k}')\cdot\textbf{k}'=Ek_0'-vk'^2y+vkk'x-pk'y+k'^2-\textbf{k}\cdot\textbf{k}'\nonumber\\
P.P'&=&EE'-\textbf{p}\cdot(\textbf{p}-\textbf{q})=M^2.
\label{dotproducts}
\end{eqnarray}
Here $\textbf{k}.\textbf{k}'=kk'[\sqrt{(1-x^2)(1-y^2)}(\cos\phi_k\cos\phi_{k'}+\sin\phi_k\sin\phi_{k'})+xy]$ with $x=\cos\theta_k$, $y=\cos\theta_{k'}$ and with the magnitude of momenta, in general, $k=|\textbf{k}|$. The tensor contractions can be splitted into the four-vector dot products as
\begin{eqnarray}
P.P_{\parallel}.K&=&\frac{(P.q_{\parallel})(K.q_{\parallel})}{q^2_{\parallel}}-P.K_{\parallel}\nonumber\\
P.P_{\parallel}.K'&=&\frac{(P.q_{\parallel})(K'.q_{\parallel})}{q^2_{\parallel}}-P.K'_{\parallel}\nonumber\\
P'.P_{\parallel}.K&=&\frac{(P'.q_{\parallel})(K.q_{\parallel})}{q^2_{\parallel}}-P'.K_{\parallel}\nonumber\\
P'.P_{\parallel}.K'&=&\frac{(P'.q_{\parallel})(K'.q_{\parallel})}{q^2_{\parallel}}-P'.K'_{\parallel}\nonumber\\
K.P_{\parallel}.K'&=&\frac{(K.q_{\parallel})(K'.q_{\parallel})}{q^2_{\parallel}}-K.K'_{\parallel}
\label{tensorcontraction}
\end{eqnarray}
The dot products in Eq.\ref{tensorcontraction} can also be simplified by taking the same assumption for the heavy quark motion as
\begin{eqnarray}
P.q_{\parallel}&=&E\omega-pq_z=E\omega-p(k'_z-k_z)=E\omega-pk'y+pkx\nonumber\\
K.q_{\parallel}&=&\omega k_0-k_zq_z=\omega k_0-k_z(k'_z-k_z)=\omega k_0-kk'xy+k^2x^2\nonumber\\
K'.q_{\parallel}&=&\omega k_0'-k'_z(k'_z-k_z)=\omega k_0'-k'^2y^2+kk'xy\nonumber\\
P'.q_{\parallel}&=&E'\omega-(p+k_z-k'_z)(k'_z-k_z)\nonumber\\
&=&E\omega-v\omega k'y+v\omega kx-pk'y+pkx+k^2x^2+k'^2y^2-2kk'xy\nonumber\\
P.K_{\parallel}&=&Ek_0-pk_z=Ek_0-pkx\nonumber\\
P.K'_{\parallel}&=&Ek_0'-pk'y\nonumber\\
P'.K_{\parallel}&=&E'k_0-(p_z+k_z-k'_z)k_z=Ek_0-vkk'y+vk^2x-pkx-k^2x^2+kk'xy\nonumber\\
P'.K'_{\parallel}&=&E'k_0'-(p+k_z-k'_z)k'_z=Ek_0'-vk'^2y+vkk'x-pk'y-kk'xy+k'^2y^2\nonumber\\
K.K'_{\parallel}&=&k_0k_0'-{k}_z{k'}_z
\label{tensordot}
\end{eqnarray}

\end{document}